\documentclass[sigconf]{acmart}

\renewcommand\footnotetextcopyrightpermission[1]{} 
\usepackage{booktabs} 
\usepackage{amsmath}
\usepackage{algorithm}
\usepackage{xcolor}
\usepackage[noend]{algpseudocode}
\usepackage{listings}
\usepackage{multirow}

 \setcopyright{none}

\begin{document}


\newcommand{\fillin}{$\rule{0.5cm}{0.15mm}$ }
\newcommand{\vecb}[1]{\boldsymbol{#1}}
\newcommand{\dv}{\boldsymbol{\hat{D}}}
\newcommand{\wv}{\boldsymbol{\hat{w}}}
\newcommand{\citen}{ \textcolor{red}{[CN]}}

\title{A Scalable Hybrid Research Paper Recommender System for Microsoft Academic}

\author{Anshul Kanakia}
\orcid{0000-0001-8332-1755}
\affiliation{%
  \institution{Microsoft Research}
  \city{Redmond}
  \state{Washington}
  \postcode{98052}
}
\email{ankana@microsoft.com}

\author{Zhihong Shen}
\orcid{0000-0000-0000-0000}
\affiliation{%
  \institution{Microsoft Research}
  \city{Redmond}
  \state{Washington}
  \postcode{98052}
}
\email{zhihosh@microsoft.com}

\author{Darrin Eide}
\orcid{0000-0000-0000-0000}
\affiliation{%
  \institution{Microsoft Research}
  \city{Redmond}
  \state{Washington}
  \postcode{98052}
}
\email{darrine@microsoft.com}

\author{Kuansan Wang}
\orcid{0000-0000-0000-0000}
\affiliation{%
  \institution{Microsoft Research}
  \city{Redmond}
  \state{Washington}
  \postcode{98052}
}
\email{kuansanw@microsoft.com}


\begin{abstract}
We present the design and methodology for the large scale hybrid paper recommender system used by Microsoft Academic. The system provides recommendations for approximately 160 million English research papers and patents. Our approach handles incomplete citation information while also alleviating the cold-start problem that often affects other recommender systems. We use the Microsoft Academic Graph (MAG), titles, and available abstracts of research papers to build a recommendation list for all documents, thereby combining co-citation and content based approaches. Tuning system parameters also allows for blending and prioritization of each approach which, in turn, allows us to balance paper novelty versus authority in recommendation results. We evaluate the generated recommendations via a user study of $40$ participants, with over $2400$ recommendation pairs graded and discuss the quality of the results using P@10 and nDCG scores. We see that there is a strong correlation between participant scores and the similarity rankings produced by our system but that additional focus needs to be put towards improving recommender precision, particularly for content based recommendations. The results of the user survey and associated analysis scripts are made available via GitHub and the recommendations produced by our system are available as part of the MAG on Azure to facilitate further research and light up novel research paper recommendation applications.
\end{abstract}


\maketitle

\section{Introduction and Related Work}

\begin{figure*}[!tb]
\includegraphics[width=\textwidth]{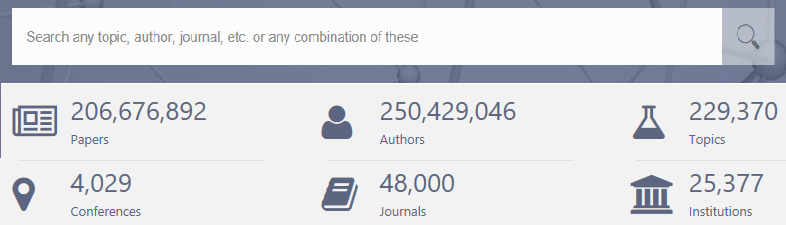}
\caption{The Microsoft Academic Graph (MAG) website preview and statistics as of October, 2018.}\label{fig:mabanner}
\end{figure*}


Microsoft Academic\footnote{https://preview.academic.microsoft.com} (MA) is a semantic search engine for academic entities \cite{sinha2015}. The top level entities of the Microsoft Academic Graph (MAG) include papers and patents, fields of study, authors, affiliations (institutions and organizations), and venues (conferences and journals), as seen in Fig.~\ref{fig:mabanner}. As of October, 2018, there are over 200 million total documents in the MA corpus, of which approximately 160 million are English papers or patents. These figures are growing rapidly \cite{dong2017}.

The focus of this article is to present the recommender system and paper similarity computation platform for English research papers developed for the MAG. We henceforth define the term `paper' to mean English papers and patents in the MAG, unless otherwise noted. Research paper recommendation is a relatively nascent field \cite{beel2016} in the broader recommender system domain but its value to the research community cannot be overstated.

The current approach to knowledge discovery is largely manual --- either following the citation graph of known papers or through human curated approaches. Additionally, live feeds such as publisher RSS feeds or manually defined news triggers are used to gain exposure to new research content. This often leads to incomplete literature exploration, particularly by novice researchers since they are almost completely reliant on what they see online or their advisors and peer networks for finding new relevant papers. Following the citation network of known papers to discover content often leads to information overload, resulting in dead ends and hours of wasted effort. The current manual approaches are not scalable, with tens of thousands of papers being published everyday and the number of papers published globally increasing exponentially \cite{dong2017, fortunato2018}. To help alleviate this problem, a number of academic search engines have started adding recommender systems in recent years \cite{bollacker1999, torres2004, barroso2006, kodakateri2009, jack2012}. Still other research groups and independent companies are actively producing tools to assist with research paper recommendations for both knowledge discovery and citation assistance \cite{gipp2009, huang2014, beel2017, bhagavatula2018}; both these use cases being more-or-less analogous.

Existing paper recommender systems suffer from a number of limitations. Besides Google Scholar, Microsoft Academic, Semantic Scholar, Web of Science, and a handful of other players, the vast majority of paper search engines are restricted to particular research domains such as the PubMed database for Medicine \& Biology, and IEEE Xplore for Engineering disciplines. As such, it is impossible for recommender systems on these field-specific search sites to suggest cross-domain recommendations. Also, a number of proposed user recommender systems employ collaborative filtering for generating user recommendations \cite{chen2018}. These systems suffer from the well known cold-start problem\footnote{Some collaborative filtering approaches assume paper authors will be system users and use citation information to indicate user intent. Even so, new authors/users still suffer from cold-start problem}. There is little to no readily available data on metrics such as user attachment rates for research paper search and recommendation sites and so it becomes difficult to evaluate the efficacy of collaborative filtering techniques without a solid active user base. Moreover, with the introduction of privacy legislation such as the General Data Protection Regulation act (GDPR) in Europe, it is becoming increasing difficult and costly to rely on user data --- which is why the Microsoft Academic website does not store personal browsing information --- making collaborative filtering all the more difficult. Finally, besides Google Scholar, the self attested paper counts of other research paper databases are in the tens of millions while the estimated number of papers from our system (as well as Google Scholar estimates) put the total number of published papers easily in the hundreds of millions. Having a tenth of the available research corpus can heavily dilute recommendations, providing incomplete and unsatisfactory results. The MA paper recommender platform aims to alleviate some of the aforementioned shortcomings by,
\begin{enumerate}
\item employing the entire MAG citation network and interdisciplinary corpus of over 200 million papers,
\item using a combination of co-citation based and content embedding based approaches to maximize recommendation coverage over the entire corpus while circumventing the cold-start problem,
\item and providing the computed paper recommendations to the broader research and engineering community so they can be analyzed and improved by other research groups.
\end{enumerate}

\begin{figure}[!htb]
\includegraphics[width=\columnwidth]{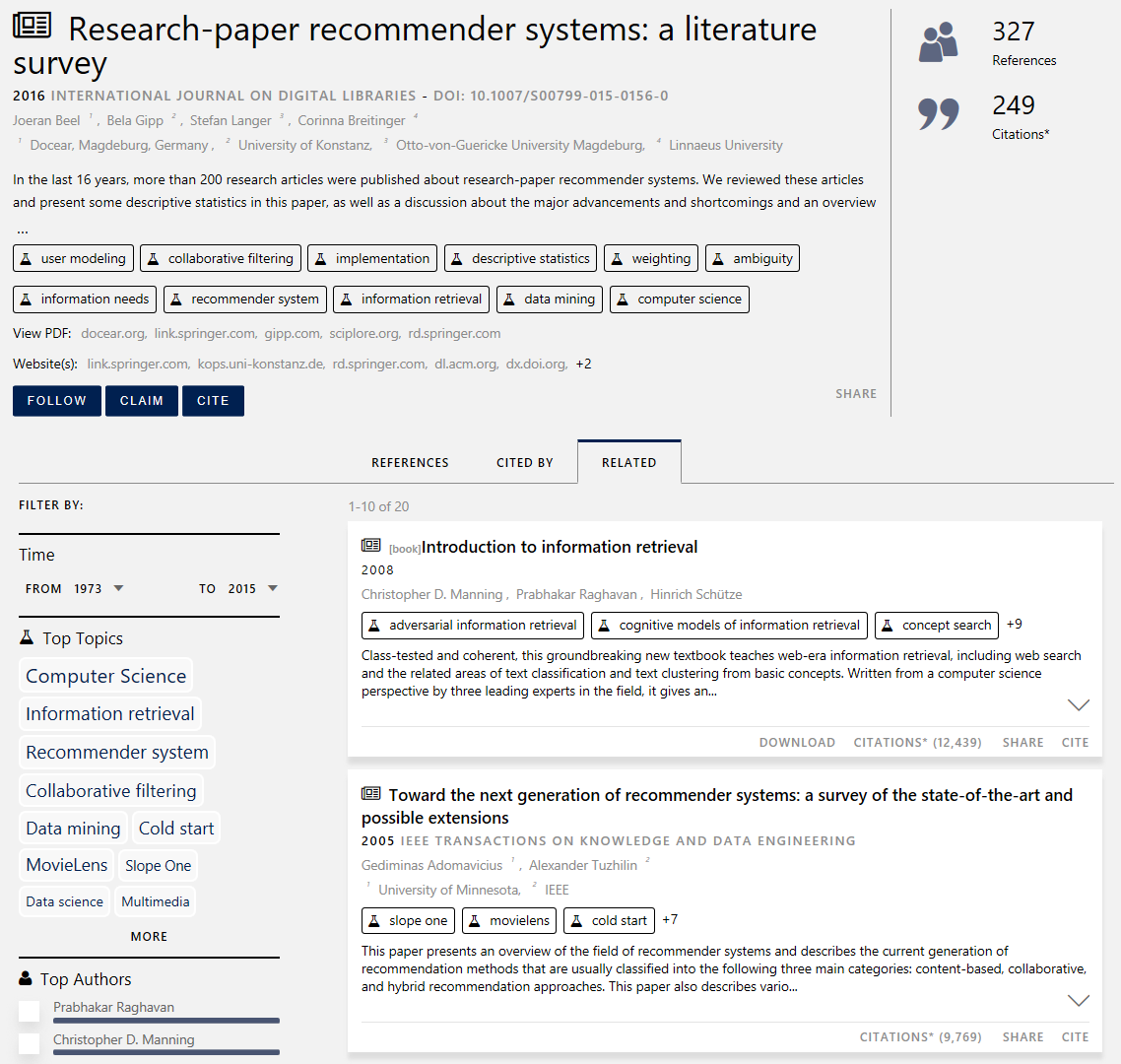}
\caption{The ``Related Papers'' tab on Microsoft Academic website paper details page with additional filters visible on the left pane.}\label{fig:ma_related}
\end{figure}

We present two possible interaction modes with the MA paper recommender platform. The first mode is via the ``Related Papers'' section on the MA search engine paper details page. Users can browse the MA search site using the novel semantic interpretation engine powering the search experience to view their desired paper \cite{sinha2015}. The paper details page, as seen in Fig.\ref{fig:ma_related}, contains a tab for browsing related papers (with dynamic filters) that is populated using the techniques mentioned here. The second mode of interaction is via Microsoft Azure Data Lake (ADL) services. The entire MAG is published under the ODC-By open data license\footnote{https://opendatacommons.org/licenses/by/1.0/} and available for use via Microsoft Azure. Users can use scripting languages such as U-SQL and python not just to access the pre-computed paper recommendations available as part of the MAG but also to generate on-the-fly paper recommendations for arbitrary text input in a performant manner. The means by which this functionality is achieved is described in the following sections.

The complexity of developing a recommendation system for MA stems from the following feature requirements:
\begin{itemize}
\item \textbf{Coverage}: Maximize recommendation coverage over the MAG corpus.
\item \textbf{Scalability}: Recommendation generation needs to be done with computation time and storage requirements in mind as the MAG ingests tens of thousands of new papers each week.
\item \textbf{Freshness}: All new papers regularly ingested by the MA data pipeline must be assigned related papers and may themselves be presented as `related' to papers already in the corpus.
\item \textbf{User Satisfaction}: There needs to be a balance between authoritative recommendations versus novel recommendations so that newer papers are discoverable without compromising the quality of recommendations.
\end{itemize}

To tackle these requirements we have developed a hybrid recommender system that uses a tunable mapping function to combine both content-based (CB) and co-citation based (CcB) recommendations to produce a final static list of recommendations for every paper in the MAG. As described in \cite{burke2002}, our approach employs a \textit{weighted mixed} hybridization approach. Our content-based approach is similar to recent work done on content embedding based citation recommendation \cite{bhagavatula2018} but differs mainly in the fact that we employ clustering techniques for additional speedup. Using purely pairwise content embedding similarity for nearest neighbor search is not viable as this is an $O(n^2)$ problem over the entire paper corpus, which in our case would be over $2.56 \times 10^{16}$ similarity computations.

\section{Recommendation Generation}\label{sec:rec_system}
The process outlined in this section describes how the paper recommendations seen on the MA website, as well as those published in the MAG Azure database, are generated. More information on the MAG Azure database is available online\footnote{https://docs.microsoft.com/en-us/academic-services/graph/} but the important thing to note is that our entire paper recommendation dataset is openly available as part of this database, if desired.

The recommender system uses a hybrid approach of CcB and CB recommendations. The CcB recommendations show a high positive correlation with user generated scores, as discussed in section \ref{sec:user_study}, and so we consider them to be of high quality. But the CcB approach suffers from low coverage since paper citation information is often difficult to acquire. To combat this issue we also use content embedding similarity based recommendations. While CB recommendations can be computationally expensive, and of lower quality than CcB recommendations they have the major advantages of freshness and coverage. Only paper metadata including titles, keywords and available abstracts are needed to generate these recommendations. Since all papers in the MAG are required to have a title, we can generate content embeddings for all English documents in the corpus, just relying on the title if need be.

The resulting recommendation lists of both approaches are finally combined using a parameterized weighting function (see section \ref{sec:combined_rec}) which allows us to directly compare each recommendation pair from either list and join both lists to get a final paper recommendation list for nearly every paper in the graph. These lists are dynamically updated as new information comes into the MAG, such as new papers, paper references or new paper metadata such as abstracts. The recommendation lists are thus kept fresh from week to week.

\subsection{Co-citation Based Recommendations}
Consider a corpus of papers $P = \{p_1, p_2, p_3\ldots p_n\}$. We use $c_{i,j} = 1$ to denote $p_i$ is citing $p_j$, $0$ otherwise. The co-citation count between $p_i$ and $p_j$ is defined as:
\begin{equation}\label{eq:cocitation} 
cc_{i,j} = \sum_{k=1}^{n} c_{i,k} c_{j,k}
\end{equation}
When $c_{i,j} \geq 1$, we call that $p_i$ is a co-citation of $p_j$ and vice-versa. Notice,  $cc_{i,j} = cc_{j,i}$. 
This method presumes that papers with higher co-citation counts are more related to each other. Alternatively if two papers are often cited together, they are more likely to be related. This approach of recommendation generation is not new, it was originally proposed in 1973 by Small et al.~\cite{small1973}. Since then others such as \cite{haruna2018} have built upon the original approach by incorporating novel similarity measures. While we stay true to the original approach in this paper, we are investigating other co-citation and network based similarity measures as future work.

CcB similarity empirically resembles human behavior when it comes to searching for similar or relevant material. Having access to paper reference information is a requirement for generating CcB recommendations. This presents a challenge since reference lists are often incomplete or just unavailable for many papers, particularly for older papers that were published in print but not digitally. Moreover, CcB recommendations are biased towards older papers with more citations by their very nature. CcB recommendations prove ineffective for new papers that do not have any citations yet and therefore cannot have any co-citations. The MAG contains complete or partial reference information for $31.19\%$ of papers, with each paper averaging approx. $20$ references. As a result, only about $32.5\%$ of papers have at least one co-citation. Nevertheless, CcB recommendations tend to be of high quality and therefore cannot be overlooked.

\subsection{Content Based Recommendations}\label{sec:content_rec}
\subsubsection{Generating paper embeddings}

CB recommendations provide a few crucial benefits to overcome the information limitations from CcB recommendations as well as the privacy concerns, system complexity, and cold-start problem inherent in other user based recommender systems approaches such as collaborative filtering. CB recommendations only require metadata about a paper that is easy to find such as its title, keywords, and abstract. We use this textual data from the MAG to generate word embeddings using the well known \textit{word2vec} library \cite{mikolov2013, mikolov2013b}. These word embeddings are then combined to form paper embeddings which can then be directly compared using established similarity metrics like cosine similarity \cite{magara2018}.

Each paper is vectorized into a paper embedding by first training word embeddings, $\vecb{w}$, using the word2vec library. The parameters used in word2vec training are provided in Table.~\ref{tab:kmeans_param} The training data for word2vec are the titles, and abstracts of all English papers in the MA corpus. At the same time, we compute the term frequency vectors for each paper (TF) as well as the inverse document frequency (IDF) for each term in the training set. A normalized linear combination of word vectors weighted by their respective TF-IDF values is used to generate a paper embedding, $\vecb{D}$. Terms in titles and keywords are weighed twice as much terms in abstracts, as seen in Eqn.~\ref{eq:doc_emb}. This approach has been applied before for CB document embedding generation, as seen in \cite{nascimento2011}, where the authors assigned a $3\times$ weight to words in paper titles compared to words in the abstract. Finally, the document embedding is normalized, $\dv = \vecb{D}/|\vecb{D}|$. Since we use cosine similarity as a measure of document relevance \cite{magara2018} embedding normalization makes the similarity computation step more efficient since the norm of each paper embedding does not need to computed every time it is compared to other papers, just the value of the dot product between embeddings is sufficient.

\begin{equation}\label{eq:doc_emb}
	\vecb{D} = 2.\hspace{-10pt} \sum_{\substack{w \in title \\ w \in keywords}} \hspace{-15pt} TFIDF(w).\wv + \hspace{-15pt} \sum_{w \in abstract} \hspace{-15pt} TFIDF(w).\wv\\
\end{equation}

\subsubsection{Clustering paper embeddings}

Our major contribution to the approach of CB recommendation is improving scalability using clustering for very large datasets. The idea to use spherical k-means clustering for clustering large text datasets is presented in \cite{dhillon2001}. The authors of \cite{dhillon2001} do a fantastic job of explaining the inherent properties of document clusters formed and make theoretical claims about concept decompositions generated using this approach. Besides the aforementioned benefits, our desire to use k-means clustering stems from the speedup it provides over traditional nearest neighbor search in the continuous paper embedding space. We utilize spherical k-means clustering \cite{hornik2012} to drastically reduce the number of pairwise similarity computations between papers when generating recommendation lists. Using trained clusters drops the cost of paper recommendation generation from a $O(n^2)$ operation in the total number of papers to a $O(n*(|c| + \lambda_c))$ operation, where $|c|$ is the trained cluster count and $\lambda_c$ is the average size of a single paper cluster. Other possible techniques commonly used in CB paper recommender system optimization, such as model classifiers \cite{bethard2010}, and trained neural networks \cite{bhagavatula2018} were investigated but ultimately rejected due to the their considerable memory and computation footprint compared to the simpler k-means clustering approach, particularly for the very large corpus size we are dealing with.

Cluster centroids are initialized with the help of the MAG topic stamping algorithm\cite{shen2018}. Papers in the MAG are stamped according to a learned topic hierarchy resulting in a directed acyclic graph of topics, such that every topic has at least one parent, with the root being the parent or ancestor of every single field of study. As of October, 2018 there are $229,370$ topics in the MAG but when the centroids for clustering were originally generated --- almost a year ago --- there were about $80,000$ topics. Of these, $23,533$ were topics with no children in the hierarchy, making them the most focused or narrow topics with minimal overlap to other fields. The topic stamping algorithm also assigns a confidence value to the topics stamped for each paper. By using papers stamped with a high confidence leaf node topic we can guess an initial cluster count as well as generate initial centroids for these clusters. Therefore, our initial centroid count $k = 23,533$. Cluster centroids are initialized by taking at most $1000$ random papers for each leaf node topic and averaging their respective paper embeddings together. Another initialization mechanism would be to take descriptive text (such as Wikipedia entries) for each leaf node topic and generate an embedding - in much the same way paper embeddings are generated using trained word embeddings - to act as an initial cluster centroid. The centroid initialization method choice is ultimately dependent on the perceived quality of the final clusters and we found that averaging paper embeddings for leaf node topics provided sufficient centroids for initializing clusters. The rest of the hyper parameters used for k-means clustering are provided in Table.~\ref{tab:kmeans_param}.

K-means clustering then progresses as usual until the clusters converge or we complete a certain number of training epochs. Cluster sizes range anywhere from $51$ papers to just over $300,000$ with $93\%$ of all papers in the MAG belonging to clusters of size $35,000$ or less. The distribution of cluster sizes is important since we do not want very large clusters to dominate computation time when generating CB recommendations. Remember that clustering reduces computation complexity of CB recommendation generation from $O(n^2)$ to $O(n*(|c| + \lambda_c))$ because each paper embedding now need only be compared to the embeddings of other papers in the same cluster. Here $|c| = k$ and $\lambda_c$ is the average cluster size. For a single paper the complexity of recommendation generation is just the second term, i.e. $|c| + \lambda_{c_i}$ where $c_i$ is the size of the cluster-$i$ that the paper belongs to so if $\lambda_{c_i}$ gets too large then it dominates the computation time. In our pipeline we found that the largest $100$ clusters ranged in size from about $40,000$ to $300,000$ and took up more than $40\%$ of the total computation time of the recommendation process. We therefore limit the cluster sizes that we generate recommendation for to $35,000$. For now, papers belonging to clusters larger than this threshold (about $7\%$ of all papers) only have CcB recommendation lists generated. In the future, we plan on investigating sub-clustering or hierarchical clustering techniques to break up the very large clusters so as to be able to generate CB recommendations for them as well. 

\begin{center}
\begin{table}[!tb]
  \begin{tabular}{lccc}
  \multirow{2}{*}{\textbf{k-means}} & init. clusters ($k$) & max iter. ($n$) & min error ($\delta$)\\
  & $23,533$ & $10$ & $10^{-3}$\\
  \midrule
  \multirow{6}{*}{\textbf{word2vec}} 
  & method & emb. size & loss fn. \\
  & \textit{skipgram} & $256$ & \textit{ns} \\
  \cmidrule(r){2-4}
  & window size & max iter. & min-count cutoff\\
  & $10$ & $10$ & $10$\\
  \cmidrule(r){2-4}
  & sample & negative & \\
  & $10^{-5}$ & 10 & \\
  \end{tabular}
  \caption{Parameters used for spherical k-means clustering and word2vec training.}\vspace{-2em}\label{tab:kmeans_param}
\end{table}
\end{center}

\vspace{-1.5em}
\subsection{Combined Recommendations}\label{sec:combined_rec}
Finally, both CcB and CB candidate sets for a paper are combined to create a unified final set of recommendations for papers in the MAG. CcB candidate sets have co-citation counts associated with each paper-recommendation pair. These co-citation counts are mapped to a score between $(0, 1)$ to make it possible to directly compare them with the CB similarity metric. The mapping function used is a modified logistic function as seen in Eq. \ref{eq:sigmoid}.
\begin{equation}\label{eq:sigmoid}
    \sigma\left(cc_{i,j}\right) = \frac{1}{1 + e^{\theta(\tau - cc_{i,j})}}
\end{equation}
$\theta$ and $\tau$ are tunable parameters for controlling the slope and offset of the logistic sigmoid, respectively. Typically, these values can be estimated using the mean and variance of the domain distribution or input distribution to this function, under the assumption that the input distribution is Gaussian-like. While the distribution of co-occurrence counts tended to be more of a Poisson-like distribution with a long tail, the majority of co-occurrence counts (the mass of the distribution) was sufficiently Gaussian-like. We settled on values $\tau = 5$ and $\theta = 0.4$ based on the mean co-occurrence count of all papers and a factor of standard deviation, respectively. In general, changing the tunable parameters of Eq.~\ref{eq:sigmoid} allows one to weigh CcB versus CB recommendations resulting in different final sets of recommendation lists that balance authoritative recommendations from CcB method with novel recommendations from the CB method. Once CcB and CB recommendation similarities are mapped to the same range $[0, 1]$ comparing them becomes trivial and generating a unified recommendation list involves ordering relevant papers from both lists just based on their similarity to the target paper. 

\section{User Study}\label{sec:user_study}
\begin{figure*}[!tb]
\includegraphics[width=\textwidth]{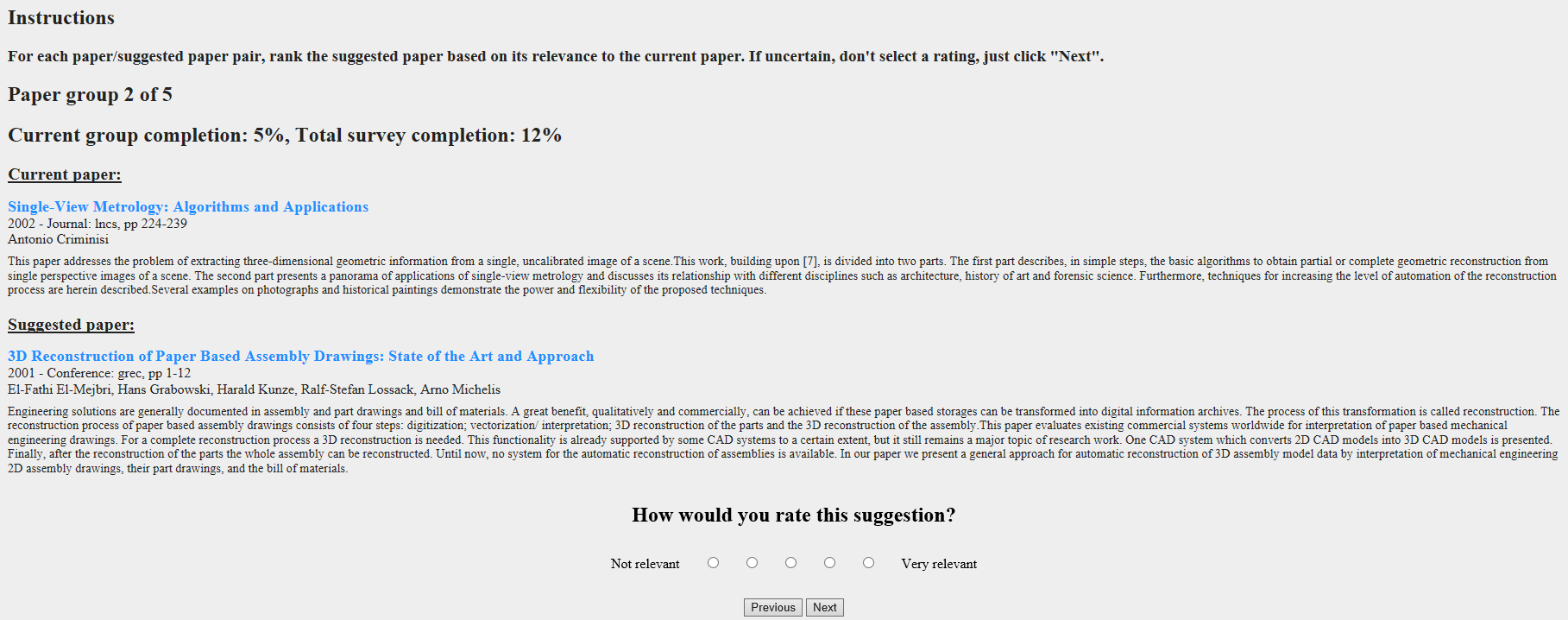}
\caption{A screenshot of the online recommender system survey.}\vspace{-1em}\label{fig:user_study}
\end{figure*}

We evaluated the results of the recommender system via an online survey. The survey was set up as follows. On each page of the study, participants were presented with a pair of papers. The top paper was one that had been identified in the MAG as being authored (or co-authored) by the survey participant while the bottom paper was a recommendation generated using the hybrid recommender platform described in the previous section. Metadata for both papers as well as hyperlinks to the paper details page on the MA website were also presented to the participant on this page. The participant was then asked to judge --- on a scale of $1$ to $5$, with $1$ being \textit{not relevant} to $5$ being \textit{very relevant} --- whether they thought the bottom paper was relevant to the top paper (See Fig.~\ref{fig:user_study}). Participants could decide to skip a survey page if they were not comfortable providing a score and carry on with the remainder of the survey.

The dataset of paper/recommended-paper pairs to show for a particular participant were generated randomly selecting at most $5$ of that participant's authored papers. This was done to ensure familiarity with the papers the participants were asked to grade, which we thought would make the survey less time consuming thereby resulting in a higher response rate. Note that while participants were guaranteed to be authors of the papers, they may not have been authors of the recommended papers. For each of a participant's $5$ papers, we then generated at most $10$ recommendations using the CcB approach, and $10$ recommendations using the CB approach. Some newer papers may have had fewer than $10$ co-citation recommendations. This resulted in each participant having to rate at most $100$ recommendations. All participants were active computer science researchers and so the survey, as a whole, was heavily biased towards rating computer science papers. We wish to extend this survey to other domains as future work since the MAG contains papers and recommendations from tens of thousands of different research domains. For now, we limited our scope to computer science due to familiarity, the ease of participant access and confidence in participant expertise in this domain.

\section{Results and Discussion}
The user survey was sent to all full-time researchers at Microsoft Research and a total of $40$ users responded to the survey, resulting in $2409$ scored recommendation pairs collected. Of these, $984$ were CcB recommendations, $15$ recommendation pairs that were both content and co-citation, and $1410$ CB recommendation pairs. The raw result dataset is available on GitHub\footnote{https://github.com/akanakia/microsoft-academic-paper-recommender-user-study}. Since at most $10$ recommendations were presented to a user using each of the two methods, we computed P@10 for CcB and CB recommendations as well as P@10 for combined recommendations. 

Since we did not include any type of explicit score normalization for participants during the survey, Table \ref{tab:user_study_stats} shows precision computed assuming, both a user score of at least $3$ as a true positive result and another row assuming a score of at least $4$ as a true positive result. Recall that users were asked to score recommended paper pairs on a scale of 1 being not relevant to 5 being most relevant. We also compute the normalized discounted cumulative gain (nDCG) for each of the three methods. Note that we use exponential gain, $(2^{score} - 1)/log_2(rank + 1)$ instead of linear gain when computing DCG.


\begin{figure}[!htb]
\includegraphics[width=\columnwidth]{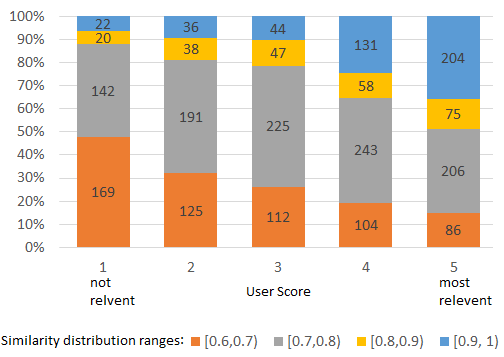}
\caption{The distribution of similarity ranges over user responses ranging from $1$ being \textit{not relevant} to $5$ being \textit{most relevant}.}\vspace{-2em}\label{fig:users_scores}
\end{figure}

\begin{center}
\begin{table}[!htb]
  \begin{tabular}{llll}
  & \textbf{CcB} & \textbf{CB} & \textbf{Combined}\\
  \cmidrule(r){2-4}
  \textbf{P@10-3} & $0.315$ & $0.226$ & $0.533$\\
  \cmidrule(r){2-4}
  \textbf{P@10-4} & $0.271$ & $0.145$ & $0.41$\\
  \cmidrule(r){2-4}
  \textbf{nDCG} & $0.851$ & $0.789$ & $0.891$
  \end{tabular}
\caption{Evaluation metrics for CcB, CB and combined recommender methods. P@10-N indicates that a user score at least N between $[1, 5]$ is considered a true positive.}\vspace{-3.5em}\label{tab:user_study_stats}
\end{table}
\end{center}

We generated Fig.~\ref{fig:users_scores} to see how the similarities computed using the MA recommender system lined up with user scores from the study. Each bar in the figure is generated by first aggregating the similarity scores for all paper recommendation pairs that were given a particular score by users. The first column is all paper recommendation pairs given a score of 1 and so on. Each column is then divided into sections by binning the paper recommendation pairs according to the similarity computed using the combined recommender method, e.g. The orange section of the leftmost bar denotes all paper recommendation pairs with a similarity between $[0.6, 0.7)$ that were given a score of 1 by the user study participants. While the absolute values of the similarity is not very important, what is important to gather from this figure is that it shows a clear positive correlation between the similarity values computed by the hybrid recommender platform and user scores. This would seem to indicate that recommendation pairs with higher computed similarity are more likely to be relevant for users, which is the desired outcome for any recommender system. This fact is reinforced by the nDCG values of each of the recommender methods. A combined nDCG of $0.891$ indicates a strong correlation between the system's computed rankings and those observed from the survey participants.

On the other hand, the observed precision values indicate that there is room for improving user satisfaction of the presented recommendations, particularly for the CB method. While we expected CcB recommendations to outperform CB results, having CB recommendations be half as precise as CcB recommendations would seem to indicate that additional effort needs to be spent in improving CB recommender quality.

\section{Conclusion}
A natural question to ask about this approach is why not use other vectorization techniques such as doc2vec\cite{le2014}, fasttext\cite{bojanowski2016} or even incorporate deep learning language models such as ULMFit\cite{howard2018}? While we settled on word2vec for the current production system, we are constantly evaluating other techniques and have this task set aside as future work. A good experiment would be to generate a family of embeddings and analyze recommendation results, perhaps with the aid of a follow-up user study to understand the impact of different document vectorization techniques on the result recommendation set. Another avenue for further research lies in tuning the weights and hyper-parameters of the recommender, such as the $\theta$ and $\tau$ parameters in the co-citation mapping function (Eq. \ref{eq:sigmoid}). We hypothesize that a reinforcement learning approach could be used to learn these parameters, given the user study as labeled ground-truth data for the training model.

In the broader scope of evaluating research paper recommender systems, there is a notable lack of literature that compares existing deployed technologies. Furthermore, there is a general lack of data on metrics such as user adoption and satisfaction, and no consensus on which approaches --- like the content and co-citation hybrid presented in this paper, or collaborative filtering, and graph analysis to name a few others --- prove the most promising in helping to tackle this problem. Part of the reason for this is that large knowledge graphs and associated recommender systems are often restricted behind paywalls, not open access or open source and hence difficult to analyze and compare. We hope to at least partly alleviate this problem by providing the entire MAG and precomputed paper recommendations under the open data license, ODC-By, so that other researchers may easily use our data and reproduce the results presented here as well as conduct their own research and analysis on our knowledge graph.

In conclusion, we presented a scalable hybrid paper recommender platform used by Microsoft Academic that used co-citation and content based recommendations in maximize coverage, scalability, freshness, and user satisfaction. We examined the quality of results produced by the system via a user study and showed a strong correlation between our system's computed similarities and user scores for pairs of paper recommendations. Finally, we made the results of our user study as well the actual recommendation lists used by MA available to researchers to analyze and help further research in research paper recommender systems. 

\bibliographystyle{ACM-Reference-Format}
\bibliography{www2019}

\end{document}